\documentclass[12pt,a4paper]{article}
\usepackage{a4wide}
\usepackage[latin1]{inputenc}
\usepackage{graphicx}
\usepackage{amsmath,amssymb}
\usepackage{natbib}
\newcommand\etal{\textit{et al.}}
\newcommand\ie{\textit{i.e.}}
\renewcommand\Re{\mathrm{Re}}
\newcommand\e{\mathrm{e}}
\newcommand\be{\begin{equation}}
\newcommand\ee[1]{\label{#1}\end{equation}}
\title{Uniform representation of the turbulent velocity profile in an open channel}
\author{Paolo Luchini\\
Universit\`a di Salerno, DIIN, 84084 Fisciano, Italy}
\begin{document}
\maketitle
\begin{abstract}
A uniform representation of the mean turbulent velocity profile in the sum of a wall function and a wake function is applied to an open channel, quantitatively determining its components. The open channel is thus found to coherently fit in to the same theoretical picture previously drawn for plane Couette, plane closed channel and circular pipe flow, and to share with them a universal law of the wall and a universal logarithmic law with a common value of von K\'arm\'an's constant.\end{abstract}
\section{Background and introduction}\label{background}
The now well established concept of Law of the Wake introduced by Coles (1956) for boundary layers extends well beyond boundary layers alone. As argued by Panton (2007) and Luchini (2018), the equally well established separation of a parallel turbulent velocity profile into a Wall Layer and a Defect Layer (going back to Millikan 1939) can be equivalently recast as a uniform representation of the mean velocity profile in the sum of a law of the wall and a law of the wake, or wall function $f(z^+)$ and wake function $G(Z)$ when one more explicitly refers to an analytical interpolation of those. In formula
\be
u^+ = f(z^+) + G(Z)
\ee{uniform}
where $u(z)$ is the mean streamwise velocity profile as a function of the wall-normal coordinate $z$, $u^+=u/u_\tau$ and $z^+=z/\ell$ are their dimensionless values in wall units, with $\ell=\nu/u_\tau$, $u_\tau=\sqrt{\tau_w/\rho}$ being the so called viscous length and shear velocity, $\tau_w$ is the wall shear stress and $\rho$, $\nu$ the fluid's density and kinematic viscosity, $Z=z/h$ is the dimensionless coordinate in outer units and $h$ the geometrical height or half-height (according as specified in the definition of $G$) of the channel.

Perhaps not equally well perceived, until recent, is that the wake function $G(Z)$ constitutes a nonnegligible contribution to the velocity profile of a parallel flow, of the same order of magnitude as it is in boundary layers, and decays relatively slowly (linearly) for $Z\rightarrow 0$. Its omission can muddy efforts to empirically determine the wall function $f(z^+)$ and the logarithmic law which describes the overlap layer. Much confusion in the literature and disagreement among the experts has, in the present author's opinion, been caused by the unverified and often taken for granted presumption that the wake function could be safely neglected. Such a presumption, on the other hand, has august precedent: the 4th edition of Schlichting (1960) states that with the logarithmic law alone ``excellent agreement is obtained not only for points near the wall but for the whole range up to the axis of the pipe" [p. 509], tantamount to saying, in more recent terminology, that the wake function $G(Z)$ is negligible for a pipe flow. In contrast, as outlined below, present data show that the wake function of pipe flow is one of the largest.

Whereas Panton (2007) extracted his representation of the wake function from a pre-assumed expression of the law of the wall, Luchini (2018) devised a method to separate the wall and wake functions without pre-assuming either, based on the difference of velocity profiles at different Reynolds numbers in the same geometry.
A lengthy examination, of all the experimental and numerical data that could be recovered at the time from the literature, led to the eventual conclusion that a single geometry-independent wall function and three individual wake functions, respectively for pipe flow, pressure-driven plane duct (closed channel) flow and turbulent Couette flow provided the best fit, in conformance with the theoretical expectation that the wall function must be universal. Analytical interpolations of such functions from Luchini (2018) are collected for convenience in Table \ref{wwtable} here.

\begin{table}
\begin{center}
\framebox{
\parbox{0.9\textwidth}{
\medskip
\ Universal wall function:
\[
f(z^+) = {{\frac{\log ( z^+ {} + 3.109 )}{ 0.392} + 4.48}} + \frac{ 7.3736 + ( 0.4930 - 0.02450  z^+ )  z^+ }{ 1 + ( 0.05736 + 0.01101  z^+ )  z^+ }  \e^{- 0.03385  z^+}
\]

\ Wake function for turbulent plane Couette flow:
\[
G(Z)=(Z-0.5)/(\exp(-25(Z-0.5))-1)
\]

\ Wake function for turbulent closed channel flow:
\[
G(Z)=\emph{Z}-0.57 Z^7
\]

\ Wake function for turbulent circular pipe flow:
\[
G(Z)=\emph{2Z}-0.67 Z^7
\]
}
}
\end{center}
\caption{Wall and wake functions for the three classical parallel geometries}
\end{table}
\label{wwtable}

In the overlap layer $\ell \ll z \ll h$ these results were consistent with the asymptotic expansion proposed by Luchini (2017), which extended the classical logarithmic law with a higher-order term in the form
\be
u^+ = \kappa^{-1}\log(z^+)+A_1g\Re_\tau^{-1}z^+ +B.
\ee{corrloglaw}
Here $A_1$ is a new universal constant and the geometry parameter $g=-hp_x/\tau_w$ is related to the hydraulic diameter and takes on the fixed values of $g=2$ for circular pipe flow, $g=1$ for turbulent closed channel flow and $g=0$ for turbulent plane Couette flow. Equation \eqref{corrloglaw} stems from the ansatz that the pressure gradient $p_x$ alone (as opposed to some other feature of the outer flow) drives the first higher term in an asymptotic expansion of $u^+$ in powers of $\Re_\tau^{-1}=\ell/h$, of which the classical logarithmic law is the leading term; it is justified by the observation that, whereas other influences from the outer region can be reasonably expected to decay when $z$ becomes smaller and smaller, the pressure gradient is constant with $z$ and keeps its value all along. Relying upon the assumption of independence from $\nu$ and $h$ (which already underpins the logarithmic law) compounded with linear dependence on $p_x$, and using dimensional analysis along similar lines to \cite{AY73} and \cite{JM07}, Luchini (2017) determined the corrected logarithmic law to be at first order linear in the wall-normal coordinate $z$, and eventually of the form \eqref{corrloglaw}. When passing to the uniform representation \eqref{uniform}, and taking $\Re_\tau^{-1}z^+=Z$, the second term of \eqref{corrloglaw} morphs into the first term of the Taylor expansion of $G(Z)$ in powers of $Z$, as it must do according to the general theory of matched asymptotic expansions, and indeed the wake functions reported in Table \ref{wwtable} contain linear terms with coefficients 0, 1, 2 for respectively plane Couette, closed channel and circular pipe flow. The best-fit values for the constants appearing in \eqref{corrloglaw}, as determined by Luchini (2017, 2018), are
\be
\kappa=0.392,\quad A_1=1,\text{ and}\quad B=4.48
\ee{constants}
respectively. The same constants also appear in the wall-function interpolation in Table \ref{wwtable}, making its asymptotic behaviour consistent with \eqref{corrloglaw}.

\begin{figure}
\centering
\includegraphics[width=0.8\textwidth]{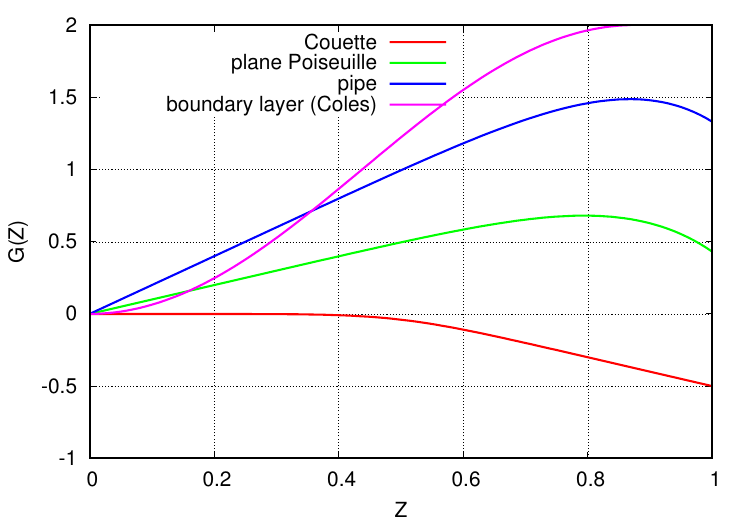}
\caption{Wake functions of different geometries as calculated in \cite{EJMB} and displayed in \cite{iTi}.}
\label{comparewakes}
\end{figure}

\begin{figure}
\centering
\includegraphics[width=0.8\textwidth]{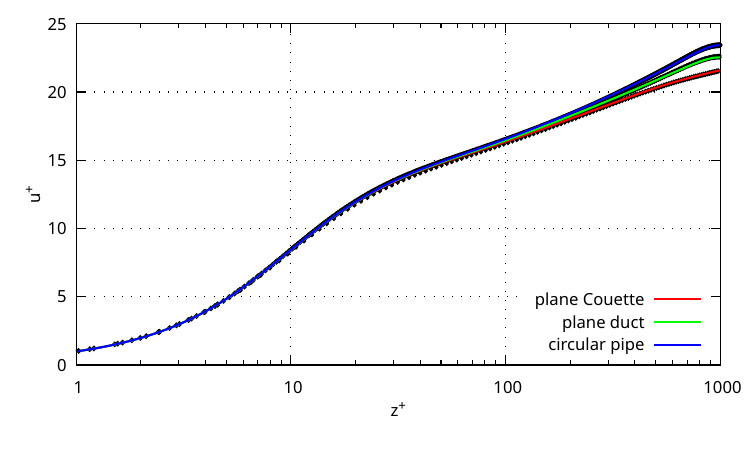}
\caption{Velocity profile versus wall-normal coordinate in wall units from \eqref{uniform} (solid lines) compared to numerical data (black dots), for three different geometries at $\Re_\tau\simeq 1000$. DNS data for the circular pipe flow are taken from \cite{Schlatter}. DNS data for the closed duct flow are taken from \cite{Moser}. DNS data for plane Couette flow are taken from \cite{Pirozzoli}.}
\label{3wakes}
\end{figure}

\begin{figure}
\centering
\includegraphics[width=0.8\textwidth]{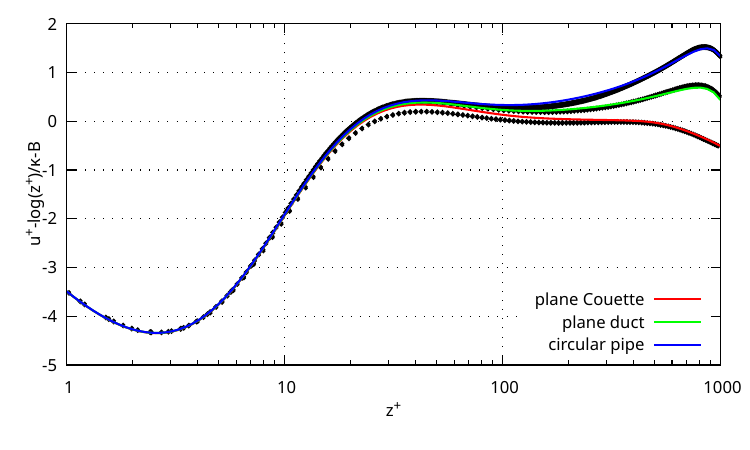}
\caption{Same data as Figure \ref{3wakes}, re-plotted as a difference from the logarithmic law.}
\label{3wakesdiff}
\end{figure}

\begin{figure}
\centering
\includegraphics[width=0.8\textwidth]{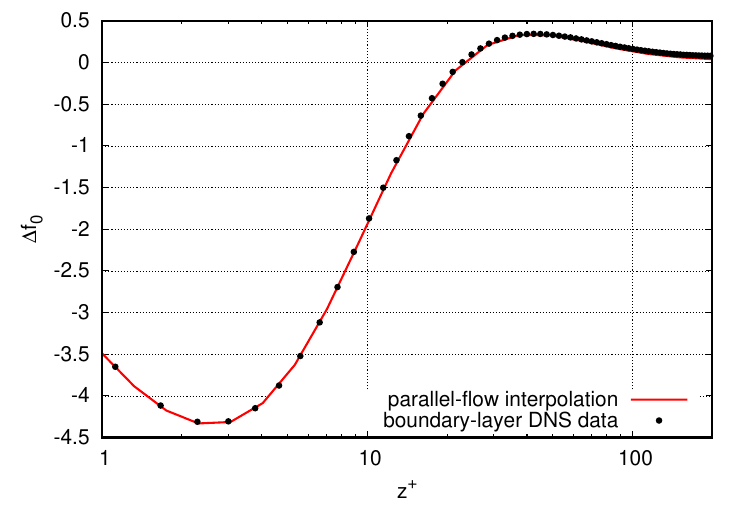}
\caption{Comparison between the velocity profile computed in the boundary-layer DNS of \cite{Jimenezbl} at $\Re_\theta=6500$ and the law of the wall from Table \ref{wwtable}, which was obtained from a fit of parallel-flow data only. Both are plotted as a difference from the logarithmic law. From Fig. 33 of \cite{EJMB}.}
\label{blayer}
\end{figure}

As laid out in Luchini (2018, 2019), the validity of the corrected logarithmic law \eqref{corrloglaw} extends for $200\ell \lesssim z \lesssim 0.5h$, and the uniform composite formula \eqref{uniform} using Table \ref{wwtable} is a valid approximation of the complete velocity profile, well within present-day experimental and numerical error, for $\Re_\tau\gtrsim 400$.
For the sake of illustration, Figure \ref{comparewakes} shows three DNS velocity profiles for plane Couette, closed channel and circular pipe flow at $\Re\simeq 1000$ compared with the corresponding predictions from Table \ref{wwtable}. Figure \ref{3wakesdiff} shows the same data again, but with the logarithmic law subtracted in order to magnify their difference.

The very good fit of velocity profiles in three different parallel geometries to \eqref{uniform} using a single wall function and wake functions that start linear in $Z$, with coefficients consistent with those geometrically derived from the corresponding pressure gradient, supports the universality of the logarithmic law and the conjecture that the pressure gradient governs its first-order correction, in addition to the general accuracy and reliability of Table \ref{wwtable}. The only visible difference to an acute eye is a downwards displacement of about -0.15 velocity wall units between the Couette numerical simulation and its uniform representation in the central part of Figure \ref{3wakesdiff}; but we have reason to believe that this difference is at least partially unreal, and can be ascribed to the numerical discretization used by Pirozzoli \etal\ (because a similar difference can be observed in the simulation of turbulent closed-channel flow by the same numerical method in Bernardini \etal\ 2014). Delving into this point here would lead us off topic, thus let us restrict ourselves to warning the reader that, as shown in \S 7 of Luchini (2018), a residual spread of order of magnitude $\pm 0.1$ is not uncommon among today's available experimental and numerical data; amid other consequences, this limits our ability to empirically extract even higher-order corrections to \eqref{uniform} and \eqref{corrloglaw}. As a pleasant surprise, as far as universality goes, the present wall-function formula a posteriori turned out to perfectly match the $0\le z^+\le 200$ range of the zero-pressure gradient boundary layer of Sillero \etal\ (2013), which it was never designed for or fitted to. See Figure \ref{blayer} adapted from Luchini (2018).% But newer geometries added to the comparison can make the case even stronger.

\section{Open channel}
Although less thoroughly studied in the literature than the three above classical geometries, an infinite open-surface channel at zero Froude and Mach numbers is another homogeneous parallel flow that lends itself well to direct numerical simulation and to the analysis techniques of Luchini (2018). Sufficiently high-Reynolds DNS data for this flow have very recently become available from Yao \etal\ (2022), who also provide a review of previous numerical and experimental explorations of the open-channel problem.

In the limit of zero Froude number (infinite gravity) the, say water-air, open surface of a channel becomes flat (with no surface waves) and, in the additional limit of infinite density and dynamic-viscosity contrast with the overlying fluid, stress-free. Its geometry can then be schematized as a doubly infinite fluid slab of height $h$ with boundary conditions $u=v=w=0$ at $z=0$ and $u_z=v_z=w=0$ at $z=h$. In laminar flow these would also be the symmetry conditions that characterize the centerline of a closed channel with a solid wall at height $z=2h$, and the open-channel problem is therefore sometimes confused with a half channel, but we must be wary that turbulent perturbations (as opposed to the mean flow) have no mirror symmetry in a true half channel, and are therefore different in the two cases. The turbulent mean velocity profiles of an open channel and a half channel are nevertheless quite similar at first sight, and only identification of the wake function can critically characterize their difference.

\begin{figure}
\centering
\includegraphics[width=0.8\textwidth]{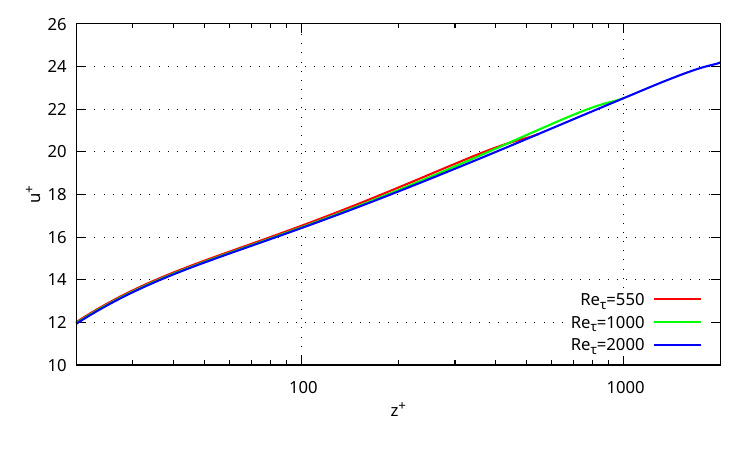}
\caption{Mean velocity profiles from Yao \etal\ (2022).}
\label{Yaoprofs}
\end{figure}

\begin{figure}
\centering
\includegraphics[width=0.8\textwidth]{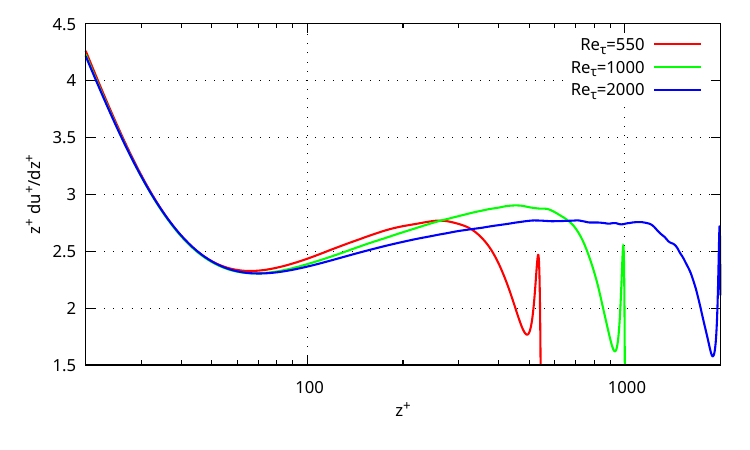}
\caption{Logarithmic derivative of the mean velocity profiles from Yao \etal\ (2022).}
\label{Yaologderiv}
\end{figure}

Yao \etal\ (2022) performed direct numerical simulations of open channel flow at $\Re_\tau=180$, $550$, $1000$, and $2000$, similar to those values at which simulations in other geometries were performed before. Their mean velocity profiles, replotted here thanks to their published data files, can be seen in Figure \ref{Yaoprofs} with the exclusion of $\Re_\tau=180$ which is too small for the present purposes.
From an interpolation of these data Yao \etal\ derive an estimate of von K\'arm\'an's constant $\kappa=0.363$, different and in fact smaller than those inferred by other authors for closed-channel flow; they remark that a region with this approximate constant value turns up in a plot of the logarithmic derivative of the velocity profile (see Figure \ref{Yaologderiv}) for $\Re_\tau=2000$ and $500\le z^+\le 1200$, and deduce that perhaps a logarithmic region has already developed for open channel flow at this Reynolds number, in contrast to the much higher Reynolds-number threshold widely believed to hold for a closed channel. In fairness, they also comment that high-order corrections to the log law were in the past introduced by \cite{AY73} and \cite{JM07} to better fit the mean velocity profile, and that the possibility of a universal $\kappa$ for open and closed channels as $\Re_\tau\rightarrow\infty$ cannot be excluded until higher-$\Re_\tau$ studies are performed. They do not seem to be specifically aware of Luchini (2018).

Starting from our previous observation that the uniform representation \eqref{uniform} and the asymptotic correction \eqref{corrloglaw} shift the asymptotic range for the validity of a logarithmic law from $\Re_\tau$ greater than some thousands or tens of thousands down to $\Re_\tau\gtrsim 400$, here we want to explore the alternate possibility that when seen in this framework Yao \etal's data, equipped with a suitable wake function unique to the open channel, are in fact compatible with the same universal wall function and universal von K\'arm\'an's constant as the other cases summarized in \S\ref{background}. If this turns out to be true we shall at the same time quantitatively determine the wake function, which can be useful in its own right to predict complete velocity profiles at all other Reynolds numbers.

\begin{figure}
\centering
\includegraphics[width=0.8\textwidth]{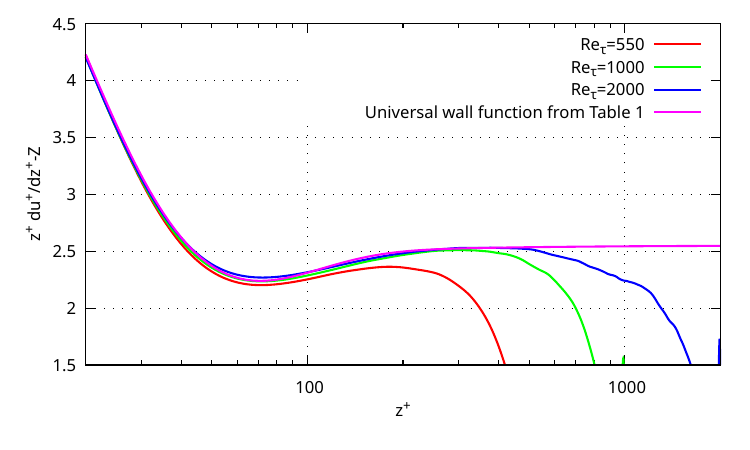}
\caption{Logarithmic derivative of the mean velocity profiles from Yao \etal\ (2022), with the linear correction \eqref{lincorr} subtracted.}
\label{subtractlin}
\end{figure}

To this end, Figure \ref{subtractlin} displays the same data as Figure \ref{Yaologderiv} with the linear correction
\be
A_1g\Re_\tau^{-1}z^+
\ee{lincorr}
(where $A_1=1$ from \eqref{constants}, and $g=1$ for an open as well as for a closed channel) subtracted. As can be seen the picture looks quite different: whereas in Figure \ref{Yaologderiv} the three profiles never really match, in Figure \ref{subtractlin} the $\Re_\tau=1000$ and $\Re_\tau=2000$ (and to a lesser degree $\Re_\tau=550$) derivative profiles match one and the same law of the wall over more than the first half of the plot. In addition, their common behaviour also matches the universal law of the wall from Table \ref{wwtable}, which is plotted on top of the same figure, including a fraction of the constant plateau that the latter develops for $z^+\ge 200$, with its universal value of $1/\kappa=2.55$. This plateau ends at $z^+\simeq 600$ for $\Re_\tau=2000$ and $z^+\simeq 300$ for $\Re_\tau=1000$, \ie\ $z^+\simeq 0.3\Re_\tau$, somewhat earlier than $z^+\simeq 0.5\Re_\tau$ which was the observed upper bound for a closed channel. In contrast what in Figure \ref{Yaologderiv} looked like a plateau for $\Re=2000$ only, has now become an oblique line which is actually part of the wake region. Most convincingly for the physical existence of the linear correction \eqref{lincorr}, the three separate ramps visible in Figure \ref{Yaologderiv} for each Reynolds number have disappeared. The top of the logarithmic region being lower also explains why $\Re_\tau=550$ falls barely out of range: from equating the beginning of the logarithmic region ($z=200\ell$ as in all other geometries) with its end ($z=0.3h$ for an open channel as opposed to $z=0.5h$ for a closed one) one derives that $\Re_\tau\gtrsim 666$ is the necessary condition for an overlap logarithmic region to exist at all, whereas it was $\Re_\tau\gtrsim 400$ for a closed channel.

\begin{figure}
\centering
\includegraphics[width=0.8\textwidth]{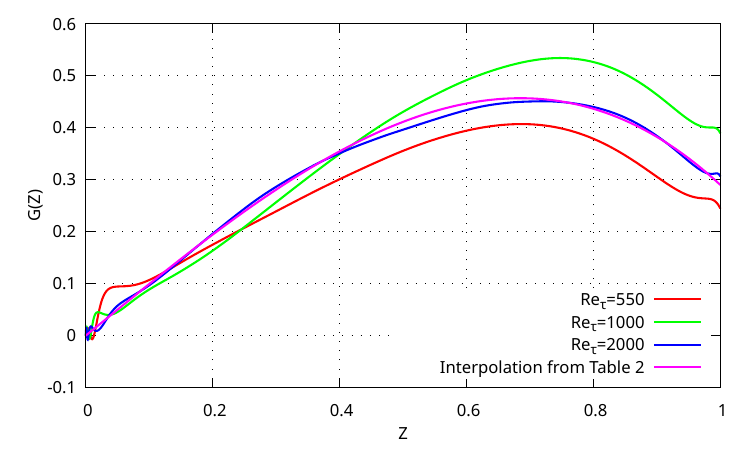}
\caption{Wake function estimates of the open channel flow at different Reynolds numbers and their analytical interpolation.}
\label{subtractwall}
\end{figure}
Once reasonable confidence is established that the velocity profiles for an open channel at multiple Reynolds numbers match the universal law of the wall from Table \ref{wwtable} in its applicable range, it remains to be seen whether the remaining part of the three velocity profiles matches one and the same wake function within acceptable error for \eqref{uniform} to hold. In order to extract the shape of the wake function from \eqref{uniform} one only needs to subtract the universal wall function from each velocity profile and plot the difference in outer coordinates. Such a plot is displayed in Figure \ref{subtractwall}, where one should note that the magnified vertical scale makes errors look bigger. The three wake function estimates sit within a $\pm 0.1$ error of each other, which is totally in range with the statistical variation observed between data of different origins for the same and/or different Reynolds numbers in all previous cases analysed by Luchini (2018). We believe that Figure \ref{subtractwall}, combined with its similarity to analogous plots formerly drawn for closed duct, plane Couette and circular pipe flow, is sufficient evidence for the existence of a Reynolds-independent, initially linear, wake function of open-channel flow. An analytical interpolation of this function is proposed in Table \ref{octable}, and superposed to the empirical estimates in Figure \ref{subtractwall}.

\begin{table}
\begin{center}
\framebox{
\parbox{0.9\textwidth}{
\medskip
\[
G(Z)=\emph{Z}-0.71 Z^{3}
\]
}
}
\end{center}
\caption{Wake-function interpolation for the open channel flow}
\end{table}
\label{octable}

\section{Conclusion}
After 100 years since its conception by Prandtl, the universality of the logarithmic law, of the law of the wall and of the law of the wake is still a matter of debate among scientists. Being such the situation it is difficult to present an absolute truth, and one must confine oneself to offering arguments in favour of one or another position. Eventually Occam's razor (the philosophical principle that ``if you have two competing ideas to explain the same phenomenon, you should prefer the simpler one") will be the only guidance. The present author in a number of papers has been furthering arguments in favour of universality, more specifically that if the wake function is properly accounted for, and is assumed to contain a linear-in-$Z$ initial term which matches (in the sense of matched asymptotic expansions) a rational asymptotic correction to the logarithmic law, mean turbulent velocity profiles not only for different Reynolds numbers but also for different geometries collapse onto a uniform representation of the form \eqref{uniform}. 

The present paper adds the open channel to the collection of geometries that fit this framework. A coherent table containing a single universal wall function and four wake functions (one for each geometry of plane closed channel, plane Couette, circular pipe and now plane open channel) is all that needed to provide a very accurate prediction of mean turbulent velocity profiles at all shear Reynolds numbers higher than a few hundred. We believe this table to be a valuable practical tool, and at the same time the simplest explanation currently available for such phenomena.

\begin{figure}
\centering
\includegraphics[width=0.8\textwidth]{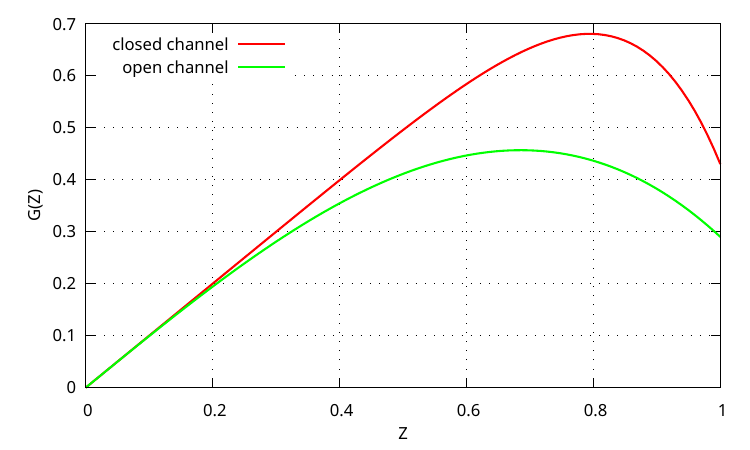}
\caption{Wake functions of open and closed channel compared.}
\label{comparewakesoc}
\end{figure}
The interpolated wake functions of an open and closed channel are compared in Figure \ref{comparewakesoc}. One can first of all observe that they are of the same order of magnitude and share the same initial slope, consistently with the general idea that these two flows behave similarly and have the same pressure gradient. The observation that they share their initial slope, in particular, is a strong supporting argument for the role of the pressure gradient proposed by Luchini (2017) in determining such slope, since other influences such as the boundary condition at the opposite wall and even the distance to the opposite wall are obviously different in the two cases and only the pressure gradient is the same. Other than this, one can observe that the wake function for the open channel is milder and peaks at a lower value that the one for a closed channel, but at the same time departs earlier from its linear behaviour, which makes for a somewhat shorter range of  the overlap layer.

\begin{figure}
\centering
\includegraphics[width=0.8\textwidth]{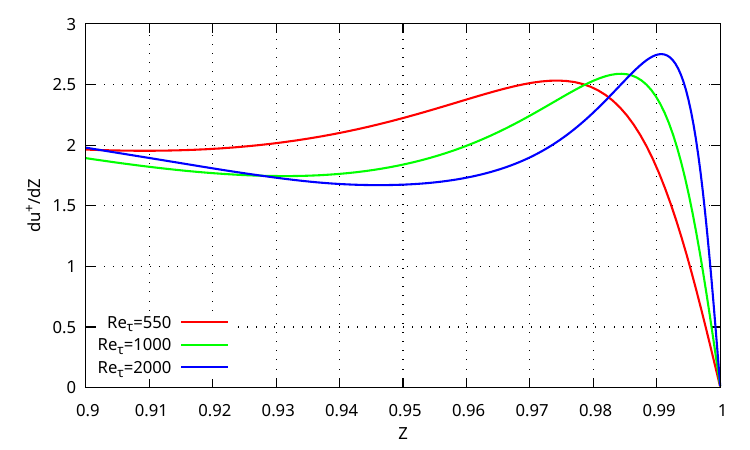}
\caption{Abrupt approach to zero of the velocity derivative near the upper free surface.}
\label{finalder}
\end{figure}
A subtler but also interesting observation is that the two wake functions exhibit different slopes at $Z=1$. This is unexpected because both the closed and the open channel rigorously have $u_z=0$ at $z=h$, and this condition clearly cannot be verified by the uniform representation \eqref{uniform} in both cases if the wake functions have different slopes. What actually happens, as can only be seen by looking at the magnified Figure \ref{finalder}, is that for the open channel $u_z$ abruptly (and non-monotonically) adjusts to zero in a thin boundary layer of thickness O$(\Re_\tau^{-1})$ near the free surface (a second ``wall layer", although of a weaker kind). As a matter of fact \eqref{uniform} is not really uniformly valid but only so outside this thin region. Physically one should be aware that the velocity derivative approaches a non-zero limit $du^+/dZ\approx 1\div 1.5$ outside this free-surface boundary layer, and in this respect closed- and open-channel flows are qualitatively different.

On the other hand the wake function is of a similar order of magnitude, and should never be neglected for an open as well as for a closed channel. For example, the open channel is not a particularly better approximation of an infinite logarithmic layer (often assumed to represent the atmospheric boundary layer) than the closed channel is. How to best simulate an infinite logarithmic layer is left as a challenge for further investigation.

\noindent \\ Declaration of Interests. The author reports no conflict of interest.
\end{document}